\documentclass[pra, amsmath, amssymb, aps, twocolumn
]
{revtex4}
\usepackage[utf8]{inputenc}
\usepackage[english]{babel}
\usepackage{graphics}
\usepackage{graphicx}
\usepackage{float}
 \usepackage{color}
 \usepackage{soul}
 \usepackage[colorlinks=true,linkcolor=blue]{hyperref}
 \usepackage{enumitem}

\usepackage{bm}

\newcommand{\tchi}{\tilde{\chi}}

\newcommand{\cR}{\mathcal{R}}

\newcommand{\rev}[1]{\textcolor{black}{#1}}

\begin{document}
\allowdisplaybreaks

\title{Localized Floquet modes in arrays of out-of-phase curved waveguides with a quasiperiodic modulation}

\author{Yaroslav V. Kartashov$^1$  and Dmitry A. Zezyulin$^2$}

\affiliation{$^1$Institute of Spectroscopy, Russian Academy of Sciences, Troitsk, Moscow, 108840, Russia\\\medskip
	$^2$School of Physics and Engineering, ITMO University, St. Petersburg 197101, Russia}

\begin{abstract}
	
We study light propagation in an array of periodically curved waveguides consisting of pairs of waveguides with out-of-phase oscillations of waveguide centers. We compute the corresponding Floquet propagation constants and find pseudocollapses where the Floquet bands shrink and, respectively, light diffraction is significantly inhibited. When, in addition, the refractive index of the waveguides in the array have quasiperiodic modulation in the transverse direction, we establish the existence of Floquet modes localized in the transverse direction and periodic in the longitudinal direction. With increase of the depth of quasiperiodic modulation of the refractive index in the array, the localized Floquet modes emerge near the pseudocollapse points of the periodic array. In array with sufficiently high frequencies of waveguide oscillations, the localized Floquet modes can exist even for weak quasiperiodic modulation which is situated below the localization transition in the array of straight waveguides.
 
\end{abstract}
\maketitle

\section{Introduction}

Contemporary optics knows several qualitatively different approaches to the localization of light. These include, in addition to conventional waveguiding due to total internal or Bragg reflections, the use of nonlinearity, localization due to periodic modulations of the waveguides in the longitudinal direction, and the addition of disorder, to name just several mechanisms. The interest in light propagation in artificially structured media, where some of these mechanisms were encountered, is enhanced by the rapid progress in waveguide  fabrication technologies, see reviews \cite{reviews,Kartashov2024}, as well as by the close analogies between the equations that describe the optical beam diffraction and the dynamics of quantum particles \cite{Longhi2009,Segev} and matter waves in condensates of ultracold atoms \cite{Carretero}. While separate localization mechanisms mentioned above are rather well studied, the combination of two or more of them may sometimes lead to unexpected dynamics of light propagation and new regimes of localization. 

The goal of this paper is to investigate the interplay between two conceptually different mechanisms of localization of light. The first mechanism involves the use of the quasiperiodic modulation of the refractive index in the transverse direction, realized by imposing modulation of waveguide depths incommensurate with periodicity of the array. Localization in such quasiperiodic structures can be achieved already in one dimension, as was theoretically predicted in \cite{AA} and confirmed in experiments with light \cite{Lahini}, atomic Bose-Einstein condensates \cite{Roati}, and cavity polaritons \cite{Goblot}. Most frequently, the localization transition in such systems occurs if the amplitude of the quasiperiodic modulation of optical potential exceeds a threshold value. The celebrated Aubry-Andr\'e model is in fact  rather special, as it possesses a peculiar property of self-duality. In a more generic case, when the self-duality is absent, localized modes are separated from the extended ones by a mobility edge in the energy spectrum \cite{Boers2007}.

The second mechanism of light localization involves periodic modulation or structuring of the waveguide array in the longitudinal direction. The majority of results in this area have been obtained for arrays of waveguides with in-phase periodic bending, where the equation describing propagation of light can be mapped to the Schr\"odingner equation for a particle confined in a periodic potential and subjected to a time-periodic ac field, using the Kramers-Henneberger transformation \cite{Longhi2009}. This correspondence enables the concept of dynamic localization of light \cite{DL}, analogous to the well-known phenomenon of localization for a wavefunction of a charged quantum-mechanical particle in external dynamical field \cite{Dunlap1986}. The dynamic localization is a resonant phenomenon which is tightly connected with the collapse of quasienergy band in the Floquet spectrum \cite{Holthaus} induced by the longitudinally periodic curvature. At the exact collapse, all propagation constants of the Floquet modes coincide, and any input beam exhibits periodic self-imaging \cite{Longhi2005}, with the period being equal to that of the longitudinal bending. As a result, the overall diffraction can be strongly suppressed. \rev{Dynamic localization, Floquet engineering,  and a variety of related phenomena can also be realized in atomic quantum gases confined in periodically driven optical lattices \cite{Eckardt}. }

In contrast to the papers mentioned above, in this work we consider the case where the centers of the adjacent waveguides oscillate out-of-phase, and hence the above simple quantum-mechanical analogy is not applicable. Subdiffractive propagation and inhibition of light tunneling in arrays with out-of-phase modulations of the refractive index in straight waveguides (different from the periodic bending) have been discussed in \cite{StalHerr,Longhi2008,Kartashov2009} and observed experimentally in \cite{Szameit2009,Torner2009}. In contrast to in-phase modulation, it has been reported that the exact quasi-band collapse is not achieved for arrays with out-of-phase refractive index modulation, even in the tight-binding approximation \cite{Longhi2008}.  \rev{At optical frequencies, out-of-phase curved waveguide arrays have been implemented in experiments with silicon  \cite{silicon} and plasmonic \cite{Sidorenko} waveguides which are characterized by appreciable losses. Low-loss waveguide arrays have been fabricated   using the femtosecond (fs) writing technique \cite{Arkhipova2023,Kartashov2024}. In the microwave range, out-of-phase curved waveguide arrays have been recently implemented in experiments with ultrathin metallic arrays of coupled corrugated waveguides \cite{micro}.  }

To study the interplay between two different localization mechanisms, we consider the situation where the out-of-phase waveguide oscillations are combined with the quasiperiodic modulation of the refractive index in the transverse direction. Several earlier papers have addressed the interplay between the Anderson localization induced by random fluctuations of the transverse distribution of the refractive index and dynamic localization in arrays with in-phase curved waveguides \cite{Borovkova}. In contrast to these studies, our system is not disordered, but features long-range order. We first revisit the case of zero quasiperiodic modulation and establish the existence of pseudocollapses in the Floquet spectrum, which correspond to a situation where the width of the Floquet band drastically contracts. Our main result consists in the existence of localized Floquet modes which emerge near the pseudocollapses of Floquet bands when the sufficiently strong quasiperiodic modulation is applied. These states are localized in the transverse direction and exhibit periodic self-imaging in the longitudinal direction. Localized Floquet modes exist within continuous intervals of oscillation amplitudes and therefore do not require the precise tuning of the parameters of the array. In particular, they can exist even for weak quasiperiodic modulation, for which the analogous array with straight waveguides would be below the localization transition.

The organization of the paper is as  follows. In Sec.~\ref{sec:model} we introduce the adopted model for the array of waveguides. Section~\ref{sec:Floquet} presents our main results regarding the Floquet modes. Section~\ref{sec:concl} concludes the paper.  

\begin{figure}
	\begin{center}	
		\includegraphics[width=0.999\columnwidth]{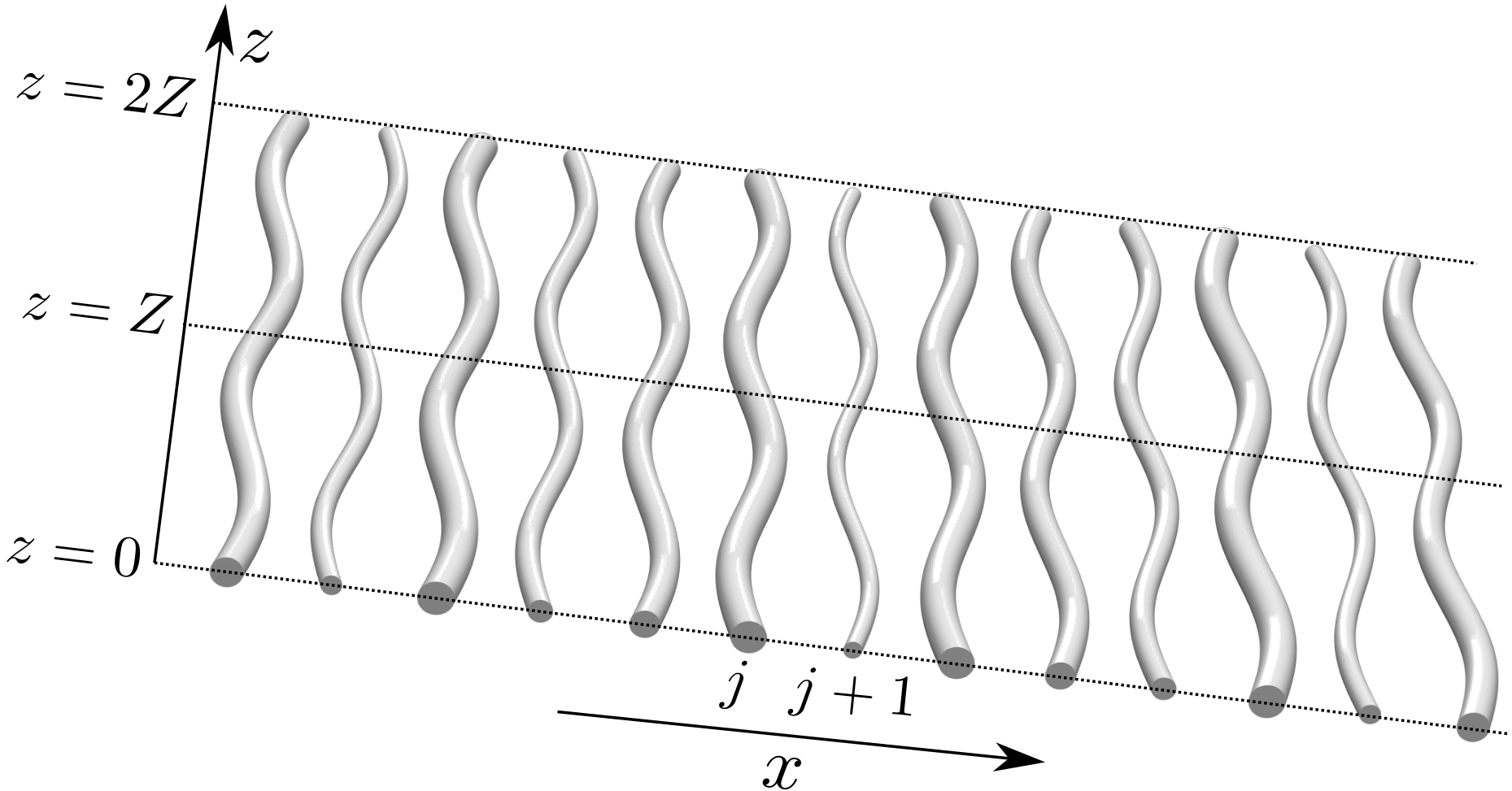}%
	\end{center}
\caption{Schematics of an array with out-of-phase oscillating waveguides with slightly different refractive indices (which are illustrated by different thicknesses of the waveguides) due to imposed transverse quasiperiodic modulation of the refractive index. Indices $j$, $j+1$, etc. enumerate the waveguides; $x$- and $z$-axes correspond to the transverse and longitudinal directions, respectively. Two longitudinal periods of the structure are shown.}
	\label{fig:array}
\end{figure}

\section{Model}
\label{sec:model}
 
We consider light propagation in an array of periodically curved waveguides, where the neighboring waveguides oscillate out-of-phase, as illustrated by the schematics in Fig.~\ref{fig:array}.  Assuming that $x$ and $z$ correspond to the transverse and longitudinal coordinates, respectively, the light field envelope $\Psi$ obeys the equation 
\begin{equation} 
i\frac{\partial \Psi}{\partial z} = -\frac{1}{2} \frac{\partial^2 \Psi}{\partial x^2} - \cR(x,z) \Psi, 
\label{eq:main}
\end{equation} 
where the function $\cR(x,z)$ describes the optical potential. Since the waveguides are periodically bent along the propagation direction, the corresponding optical potential is periodic as well,  $\cR(x, z) = \cR(x, z+ Z)$, where $Z$ is the period of the longitudinal oscillations. The array of out-of-phase oscillating waveguides is described by:
\begin{eqnarray}
\label{eq:potential}
\cR(x, z) = \sum_j p_j e^{-[x - jd - (-1)^j r\sin(\omega z)]^2/a^2},
\end{eqnarray}
where integer $j$ enumerates the waveguides, $\omega=2\pi/Z$ is the frequency of longitudinal modulation,  $d$ is the mean distance between the waveguides, $a$ is  the effective width of the waveguides, and $r$ is the amplitude of waveguide oscillations. The coefficient $p_j$ describes the refractive index modulation depth in $j$-th waveguide. We consider the array of waveguides with quasiperiodic modulation in the transverse direction. Respectively, the waveguide depths  $p_j$ are distributed according to the following law (which is similar to  the Aubry-Andr\'e (AA) model \cite{AA}) 
\begin{equation}
p_j = p^{0}[ 1 + \delta \cos(2\pi \varphi j + \theta)].
\end{equation}
Here $p^{0}$ is the mean value, $\varphi$ is an irrational number, which defines refractive index modulation incommensurate with waveguide spacing,  $\delta \in [0,1)$ is the relative depth of the quasiperiodic modulation; $\theta \in [0, 2\pi)$ is the phase shift.

Considering the experiments with modulated arrays written in fused silica slabs by a fs-laser \cite{Szameit2009,Arkhipova2023,Kartashov2024}, we adopt the dimensionless values $d=3$ and  $a=0.5$. For operating wavelength of $800$~nm, these values correspond to waveguide spacing $d=30~\mu$m and width $a=5~\mu$m, respectively; the representative dimensionless waveguide depth $p^0=4.5$ corresponds to the refractive index contrast $\Delta n \approx 5.0\times 10^{-4}$. Since the longitudinal oscillations of any pair of adjacent waveguides are out-of-phase, we limit the amplitude of oscillations $r$ to the interval $r\in [0, 1]$ to avoid overlap of the waveguides. The representative longitudinal oscillation frequency $\omega=0.2$ corresponds to oscillation period $Z=10\pi$ that in physical units corresponds to $\approx 33\ $mm.

To eliminate the boundary effects that are not relevant to the current study, we substitute the irrational number $\varphi$ with its rational approximant, $\varphi \approx  M/N$, where $M$ and $N$ are large, coprime, odd and even integers, respectively. Then the optical landscape corresponding to $\cR(x,z)$ becomes periodic in the transverse direction: $\cR(x, z) = \cR(x+N d, z)$. Hence the finite array of $N$ waveguides is naturally suited for periodic boundary conditions. The accuracy of such rational approximation can be controlled by increasing $N$ and $M$ \cite{RA,Yang2024}.

\begin{figure}
	\begin{center}	
				\includegraphics[width=0.999\columnwidth]{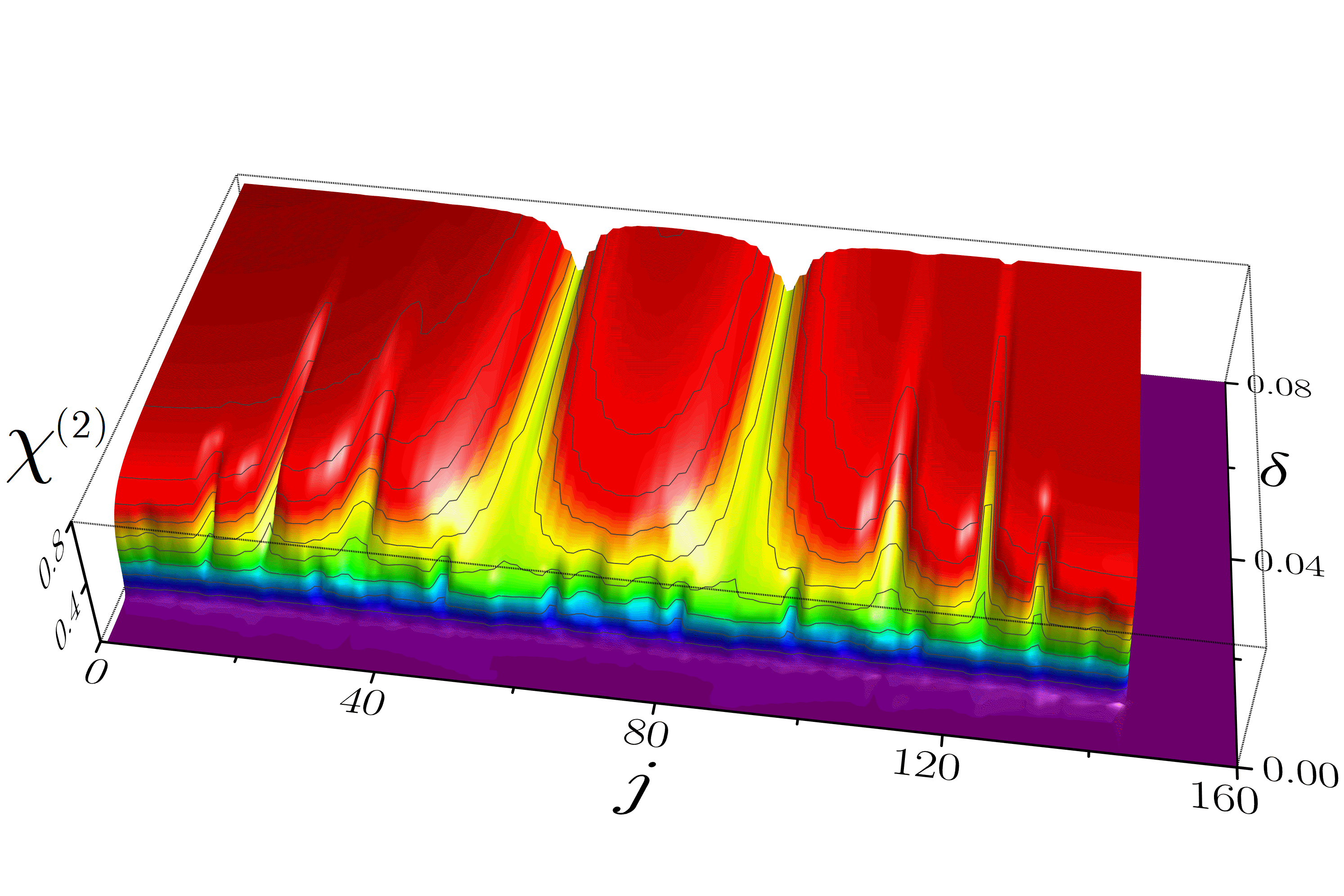}%
	\end{center}
\caption{Form-factors $\chi^{(2)}$ of guided modes of the array of straight waveguides with $r=0$, for different strengths of the quasiperiodic modulation $\delta$. The plot includes form-factors for $160$ modes with the largest propagation constants. In this figure and below, we use average waveguide depth $p^0=4.5$, the rational approximant $\varphi \approx 233/144$ of true irrational number $\varphi = (5^{1/2}+1)/2$, and phase shift $\theta \approx 1.1$.}
	\label{fig:LDT}
\end{figure} 

\rev{In the tight-binding approximation (that becomes accurate for sufficiently deep optical potentials with $p^0 \gg 1$), the envelope equation (\ref{eq:main}) with the optical potential given by Eq.~(\ref{eq:potential}) can be transformed into a lattice model governed by the AA Hamiltonian with out-of-phase oscillations in the hopping amplitudes between neighboring sites:
\begin{eqnarray}
    i \frac{dA_j(z)}{dz}   = \frac{C^0}{2}[1 - (-1)^j \rho \sin(\omega z)] A_{j-1} \nonumber \\[2mm] + \frac{C^0}{2}[1 +  (-1)^j \rho \sin(\omega z)] A_{j+1}  
    + V^0 \cos(2\pi \varphi j + \theta) A_j,  
\end{eqnarray}
where $j=0, \pm 1, \ldots$,  $C^0>0$ is the mean hopping amplitude, $\rho\in [0, 1]$ is the relative strength of out-of-phase oscillations, and $V^0>0$ is the strength of the quasiperiodic potential. When the periodic driving along the propagation distance $z$ is absent, the AA model is self-dual \cite{AA}, and its eigenstates, $A_j(z) = a_j e^{i b z}$, are delocalized (localized) for $V^0 <(>) 2C^0$. AA Hamiltonians with an in-phase periodic oscillations of the hopping amplitudes feature qualitatively distinct properties and, in particular, can be characterized by the presence of the mobility edge separating localized and delocalized eigenstates coexisting in the Floquet spectrum \cite{Sarkar}. To the best of our knowledge, the tight-binding AA model with out-of-phase oscillations in hopping amplitudes has not been studied yet. 
}

To better understand the effect of waveguide oscillations on light localization in this system, we first review the properties of eigenmodes of the array of straight waveguides with $r = 0$. We compute the spectrum of stationary guided modes $\Psi_j = \psi_j(x) e^{ib_jz}$, where $b_j$ are the propagation constants enumerated in the descending order: $b_1 > b_2 >\ldots$. For each stationary mode, we characterize its localization along the transverse direction using the form-factor defined as 
\begin{equation}
\rev{ 
\chi^{(q)} =   U^{-q} \int_{0}^{L}  |\psi(x)|^{2q} dx}, \qquad q>0,
\end{equation} 
where $L=Nd$ is the width of the array, and $U= \int_{0}^{L}  |\psi(x)|^2dx$ is the mode power. \rev{In the physics of Anderson transitions \cite{Evers}, quantities $\chi^{(q)}$ are widely known as (generalized) inverse participation ratios (IPRs). For large number of waveguides $N\gg 1$ (or, equivalently, for large system size $L\gg 1$) the generalized IPRs have scaling $\chi^{(q)} \propto N^{-D(q-1)}$, where $D=0$ for localized states and $D=1$ for delocalized states.}

A representative plot showing form-factors  $\chi^{(2)}$  of different eigenmodes versus strength of the quasiperiodic modulation $\delta$ is shown in Fig.~\ref{fig:LDT}. One can see that there is a sharp transition between delocalized and localized modes which is achieved as the modulation strength $\delta$ increases. The localization transition occurs for exactly $N$ modes with the largest propagation constants, while the rest of the spectrum contains delocalized modes for any value $\delta$. \rev{Thus, for sufficiently large $\delta$, the spectrum of array of straight waveguides has a mobility edge which separates 144 modes with large propagation constants $b$ and large form-factors $\chi^{(2)}$ from the rest of delocalized modes with much smaller form-factors.}

\section{Floquet modes}
\label{sec:Floquet}
 
\subsection{Computation of the Floquet modes} 

Further, we proceed to the array with oscillating waveguides and compute the Floquet modes induced by the periodic longitudinal variation of the structure. To this end, we assume that, up to a suitable accuracy, the light propagation in the waveguide can be completely described in terms of $N$ modes with largest propagation constants and the excitation of other modes does not occur. The validity of this assumption is discussed below. Let $\psi_1^0(x), \ldots, \psi_N^0(x)$ be orthonormal eigenvectors corresponding to $N$ modes of the array at $z=0$, having the largest propagation constants. We simulate propagation of each input vector $\psi_j^0(x)$ up to $z=Z$ and denote the obtained fields as $\psi_j^1(x)$. Then we create the matrix of projections $M = (  m_{j,k} )$, where $m_{j,k} =   \int_{0}^{L} (\psi_j^0)^* \psi_k^1 dx$, and the asterisk means complex conjugation. The matrix $M$ is similar to the monodromy matrix in the  Floquet theory \cite{Yakubovich}. In particular, its eigenvalues are the so-called Floquet multipliers $\rho_j = e^{i\beta_j Z}$, and $\beta_j$ can be referred to as Floquet propagation constants (using the quantum-mechanical analogy, the values $-\beta_j$ correspond to quasienergies \cite{Holthaus}). In the standard Floquet theory, the multipliers are unimodular, i.e., $|\rho_j|=1$, and, respectively, the characteristic exponents are purely real: $\mathrm{Im}\, \beta_j = 0$. However, in our case, these properties hold only approximately, as small radiation is always present and light very slowly leaks into the modes that have been truncated in the computation. Therefore the Floquet propagation constants may have small positive imaginary parts which correspond to the energy dissipation. For all calculations presented below, the imaginary parts of all Floquet propagation constants do not exceed $10^{-3}$, and hence the light tunneling to the truncated modes can be safely neglected.

The eigenvectors of the matrix of projections $M$ can be used to create the input light distributions corresponding to the Floquet modes. Let $V = (v_{j,k})$ be the matrix composed of eigenvectors of $M$, and the initial distribution is prepared as 
\begin{equation}
\Psi^F_j(x,z=0) =  \sum_{k=1}^N v_{j,k} \psi_k^0(x), \quad j=1,2,\ldots, N,
\end{equation}
then the evolution of this field is given as $\Psi^F_j(x,z) = e^{ib_jz}u_j(x,z)$, where $u_j(x,z) = u_j(x, z+Z)$. Hence each Floquet mode periodically reproduces its transverse intensity distribution along the propagation distance, i.e., $|\Psi^F_j(x,z)| = |\Psi^F_j(x,z+Z)|$, as long as the imaginary part of   Floquet propagation constant $\beta_j$ is negligible. 

\begin{figure}
	\begin{center}	
		\includegraphics[width=\columnwidth]{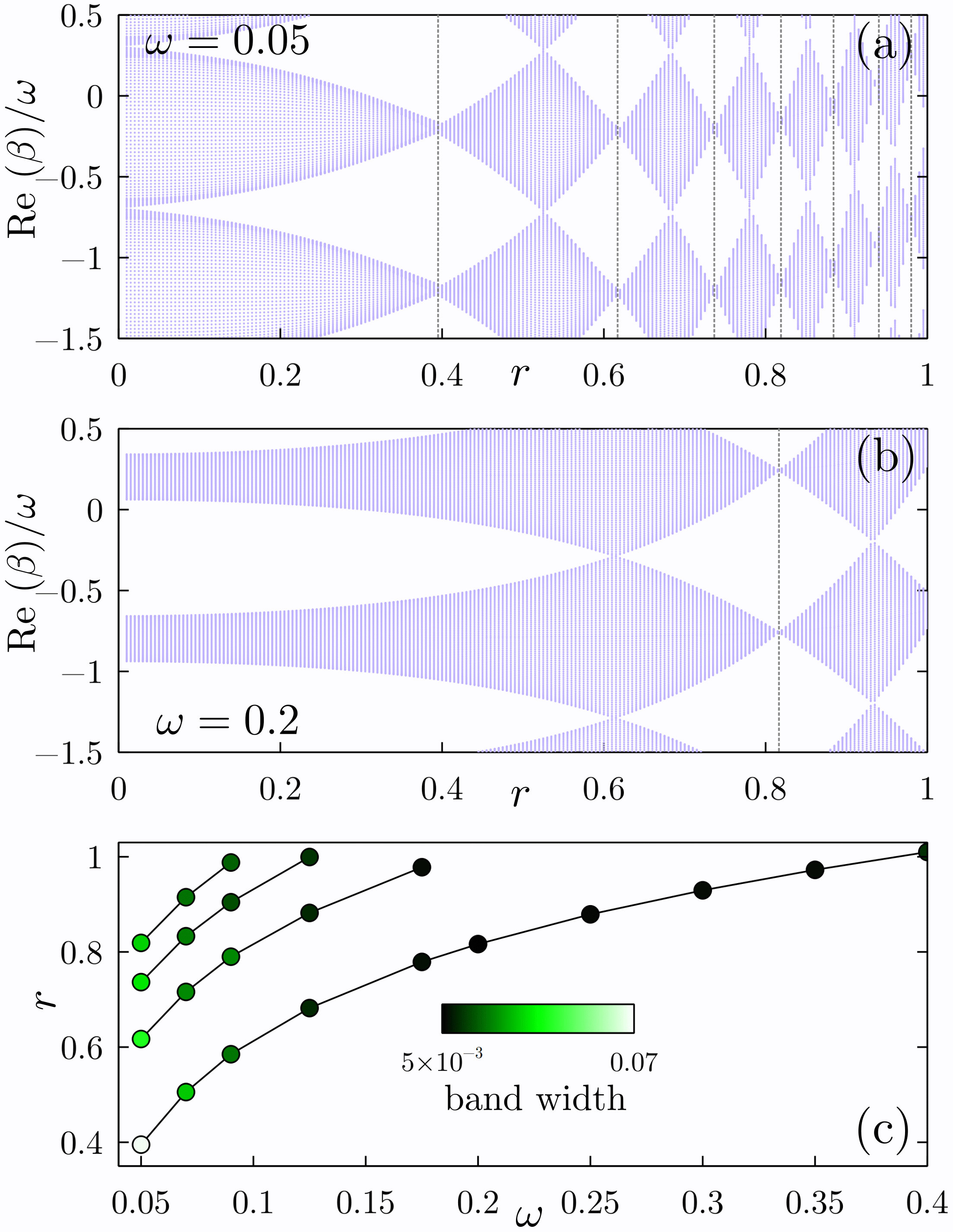}%
	\end{center}
\caption{Real parts of Floquet propagation constants of eigenmodes of oscillating array for zero quasiperiodic modulation ($\delta=0$) and for increasing amplitude of longitudinal oscillations $r$ at $\omega=0.05$ (a) and $\omega=0.2$ (b). Each panel shows two replicas of the Floquet spectrum (two longitudinal Brillouin zones). Panel (c) shows the amplitudes $r^*$ at which Floquet band achieves a pseudocollapse (i.e. its width becomes minimal). The amplitudes corresponding to first four pseudocollapses are shown. Circles are obtained from the calculation, and solid lines are guides for an eye. Colors of the circles indicate the width of the band in each pseudocollapse point.}
	\label{fig:delta=0}
\end{figure} 

\begin{figure}
	\begin{center}
		\includegraphics[width=0.999\columnwidth]{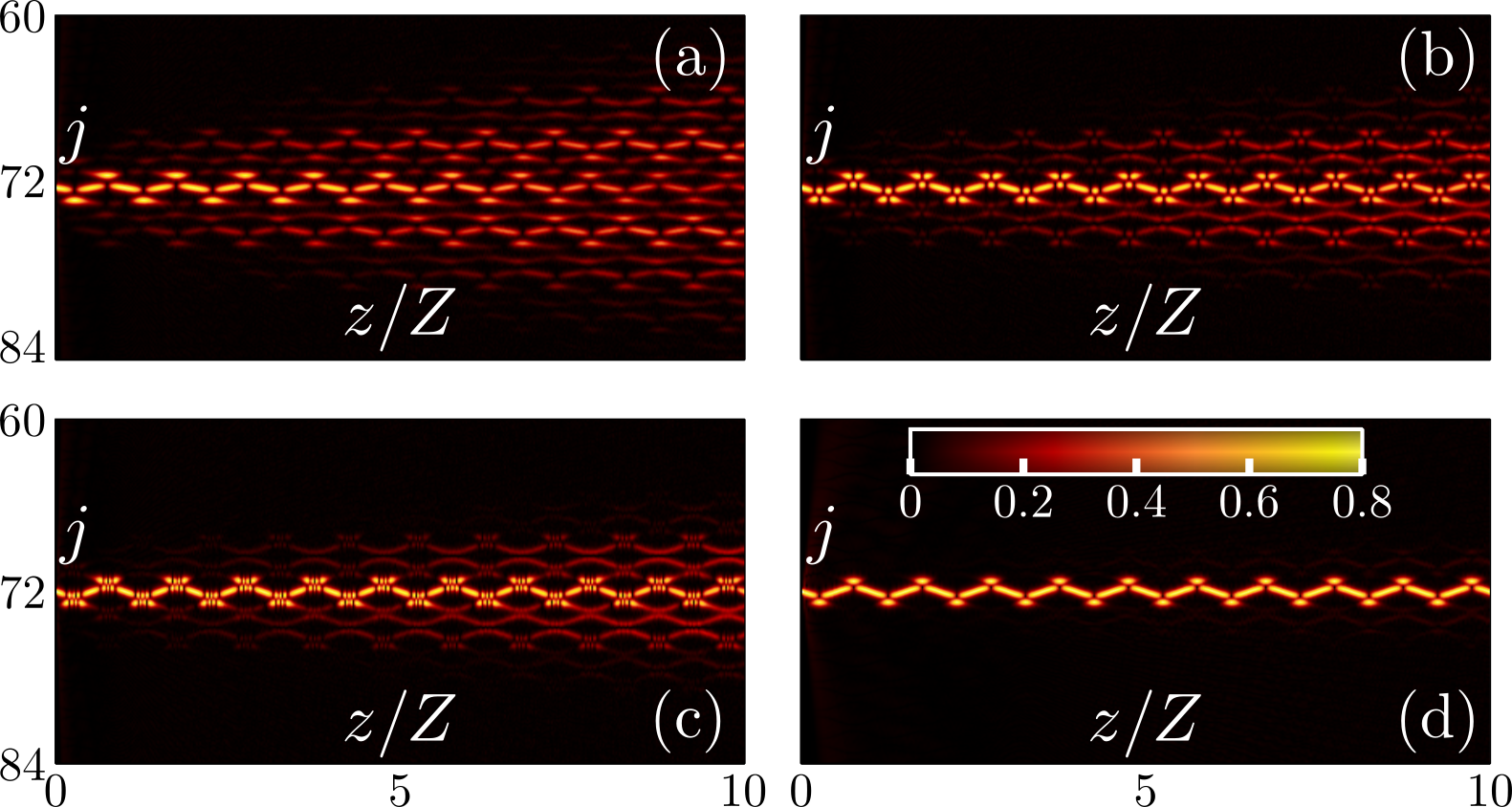}
	\end{center}
\caption{Propagation of an input field corresponding to the single-waveguide excitation for amplitude of oscillations $r$ corresponding to three pseudocollapses at $\omega=0.05$ [$r=0.395, 0.617, 0.819$ in (a)-(c)] and one pseudocollapse at $\omega=0.2$ [$r=0.817$ in (d)]. Only a small part of the waveguide array is shown in each figure. The propagation distance corresponds to ten longitudinal periods $Z$ in each figure. \rev{The colorbar shows $|\Psi|$ and  applies to all plots.}}
	\label{fig:periodic_single}
\end{figure}

\begin{figure*}
	\begin{center}	
		\includegraphics[width=0.8\textwidth]{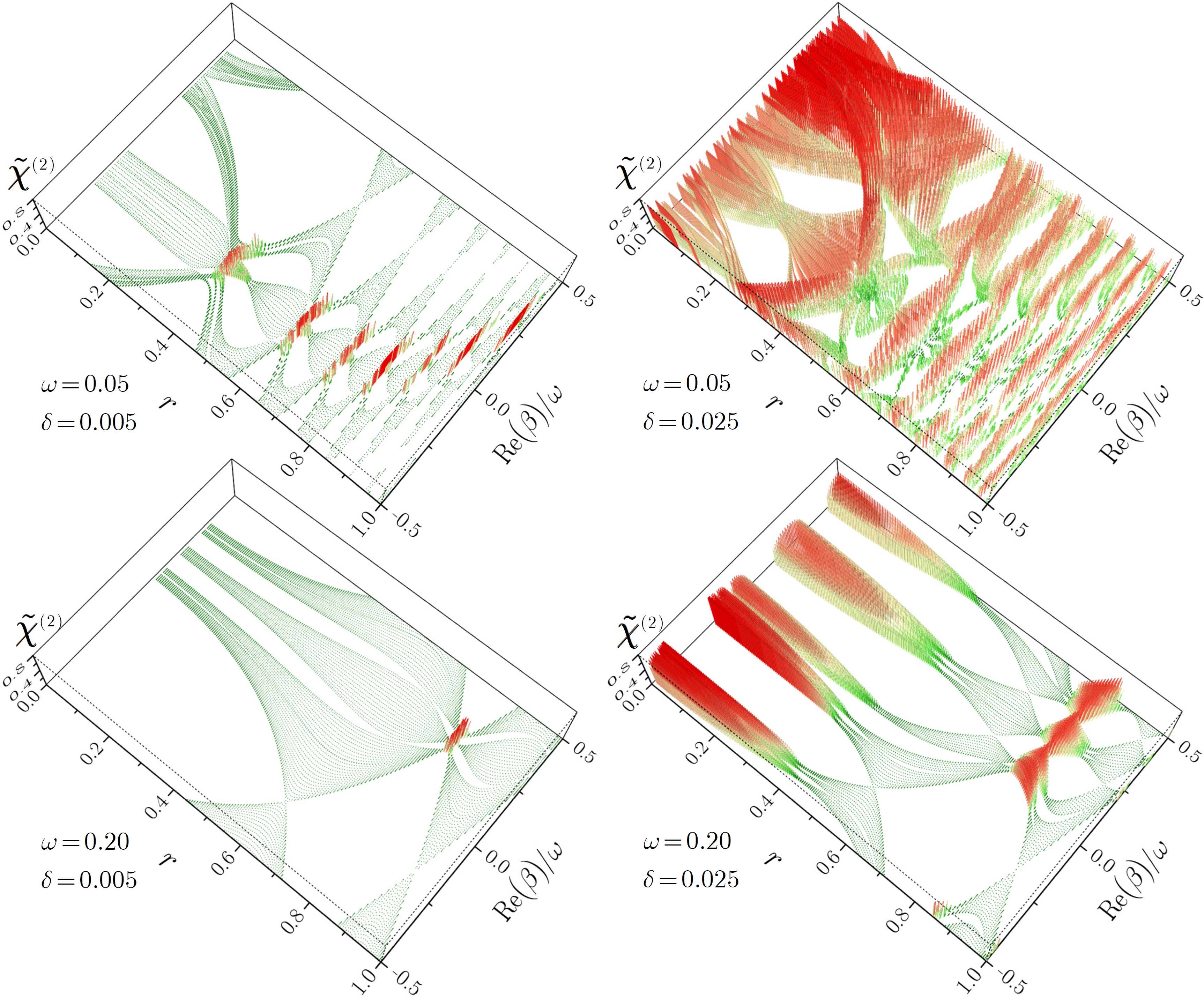}%
	\end{center}
\caption{Real parts of Floquet propagation constants $\beta$ and averaged form-factors $\tchi^{(2)}$ versus amplitude of waveguide oscillations $r$ for Floquet modes in arrays with two different frequencies of the longitudinal oscillations $\omega$ and two different strengths of the quasiperiodic modulation $\delta$. Specific values of $\omega$ and $\delta$ are placed next to the corresponding plots.   }
	\label{fig:delta}
\end{figure*}

\begin{figure}
	\begin{center}
		\includegraphics[width=0.999\columnwidth]{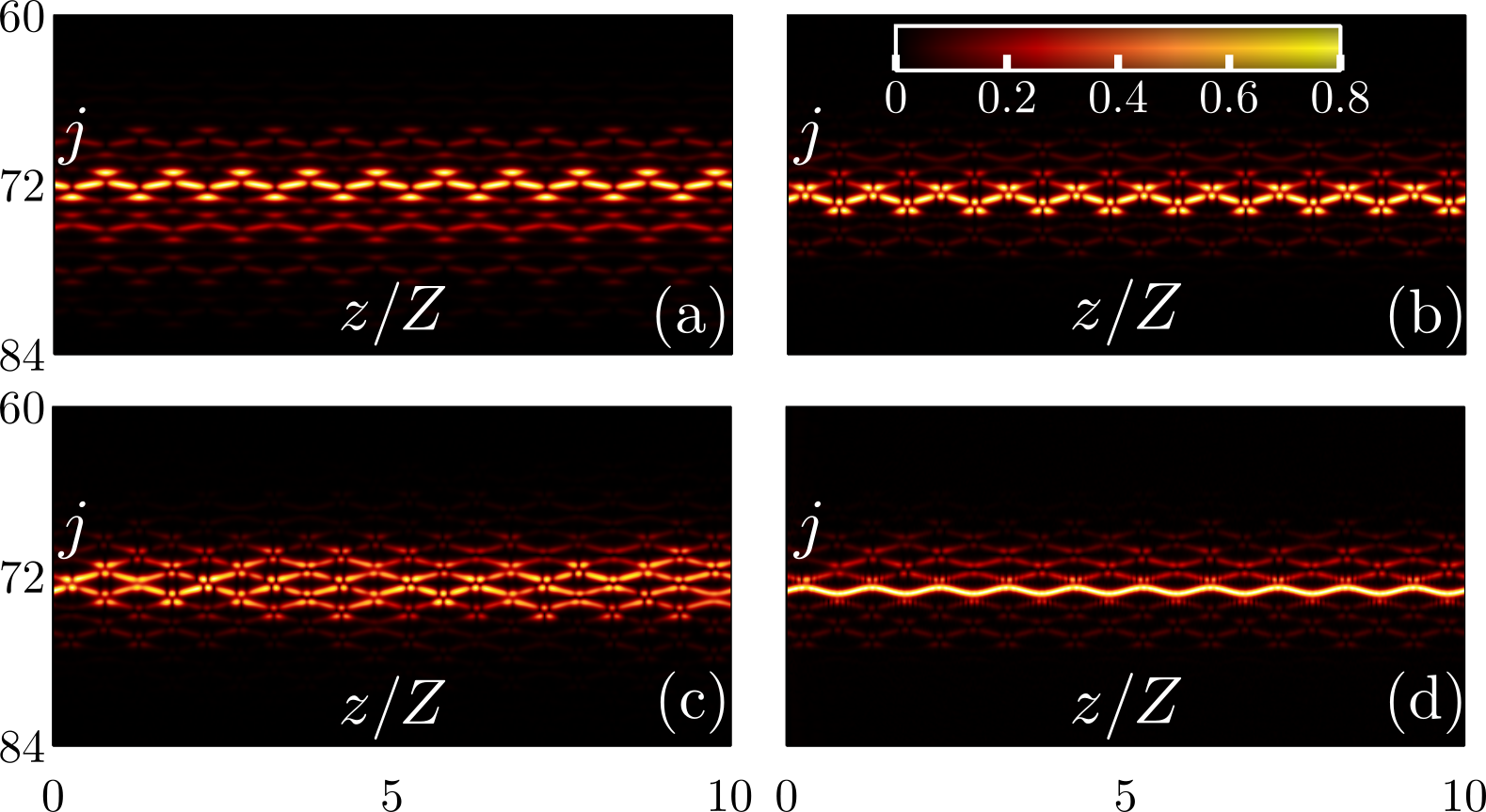}%
	\end{center}
\caption{Propagation of representative localized Floquet eigenmodes in arrays with quasiperiodic modulation strength $\delta=0.005$, oscillation frequency $\omega=0.05$ and amplitudes $r=0.395$ (a) and $r=0.617$ (b) (corresponding to pseudocollapse of the band in arrays without quasiperiodic transverse modulation). \rev{  Propagation of the initial distribution corresponding to the Floquet mode in (b) in media with relatively weak (c) and strong (d) self-focusing nonlinearity.  The colorbar shows $|\Psi|$ and applies to all plots.}}
	\label{fig:floquet}
\end{figure}   

\subsection{Pseudocollapses in the array without quasiperiodic transverse modulation} 

In Fig.~\ref{fig:delta=0}(a,b) we show real parts of Floquet propagation constants versus increasing amplitude of the waveguide oscillations $r$, for two different frequencies $\omega$ in the array without transverse quasiperiodic modulation, i.e. at $\delta =0$. The Floquet propagation constants are defined up to an integer multiple of the frequency $\omega$, and hence the Floquet spectra are periodic along the vertical axis. Thus, in Fig.~\ref{fig:delta=0}(a,b) we show two replicas of the Floquet spectra for each frequency $\omega$. The most prominent feature visible in Fig.~\ref{fig:delta=0}(a,b) is that the Floquet bands drastically collapse at certain values of the amplitude of waveguide oscillations. This behavior resembles the Floquet band collapses which are known to occur in arrays with \textit{in-phase} waveguide oscillations and lead to dynamic localization. However, in our case the collapses are not perfect and the band width remains finite. Hence in what follows, we use the term \textit{pseudocollapse} to refer to the situation where the Floquet band width attains a minimum (see also Ref.~\cite{Longhi2008}, where this term was used in a similar context). Respectively, the pseudocollapses are characterized by nonzero Floquet band widths plotted in Fig.~\ref{fig:delta=0}(c) for four pseudocollapses corresponding to small amplitudes of the longitudinal deformation $r$. Figure~\ref{fig:delta=0}(c) indicated that the pseudocollapses are far from being perfect for small oscillation frequencies $\omega$. However, the minimal band width in pseudocollapse point notably decreases at larger frequencies. Additionally, the amplitudes $r$, where the pseudocollapses are achieved, increase with $\omega$. Due to this, within selected interval of $r$ values one observes several pseudocollapses at small $\omega=0.05$ [see Fig.~\ref{fig:delta=0}(a)], whereas for sufficiently fast oscillations $\omega=0.2$ there is a single pseudocollapse point within this range of amplitudes $r$ [see Fig.~\ref{fig:delta=0}(b)].

\rev{We notice that for each pseudocollapse in Fig.~\ref{fig:delta=0}(c)  the  amplitude $r^*$ is an increasing function of the frequency $\omega$. This behavior is drastically different from the analogous dependencies in arrays of in-phase curved waveguides \cite{Longhi2005}, where the  amplitude of the curvature modulation ($r$) corresponding to the quasienergy collapse is inversely proportional to the frequency $\omega$.}

In the array without the quasiperiodic modulation, the Floquet modes are generically extended in the transverse direction. At the same time, at the point where the perfect Floquet band collapse is achieved, any input field distribution must be periodic along the propagation distance (because it excites Floquet modes that all have the same Floquet propagation constants in the perfect collapse point), and hence the diffraction is suppressed in a sense that the field distribution is exactly reproduced after each longitudinal period $Z$. In our case, under not ideal band collapse conditions, the diffraction remains, but it becomes strongly suppressed, especially when the band width in pseudocollapse point substantially decreases. This is illustrated in Fig.~\ref{fig:periodic_single} (compare dynamics at $\omega=0.05$ and $\omega=0.2$), where the input field corresponds to the excitation of a single waveguide situated in the center of the array.

\subsection{Localized Floquet modes in the quasiperiodically modulated array}

Further, we proceed to the  nonzero transverse quasiperiodic modulation in the oscillating array. In Fig.~\ref{fig:delta} we show real parts of Floquet propagation constants and averaged form-factors of Floquet modes versus amplitude of waveguide oscillations $r$ for two different strengths of the quasiperiodic modulation, $\delta=0.005$ and $\delta=0.025$, and two different oscillation frequencies, $\omega=0.05$ and $\omega=0.2$. The plotted form-factors $\tchi^{(q)}$ are averaged over one longitudinal period $Z$, i.e., 
\begin{equation}
\tchi^{(q)} =  Z^{-1} (U^F)^{-q} \int_0^Z dz  \int_{0}^{L}  |\Psi^F(x,z)|^{2q}dx,
\end{equation}
where $U^F = \int_{0}^{L}  |\Psi^F(x,z)|^2dx$ is the  power of the Floquet mode (which does not depend on $z$ in view of the power conservation).

Two different quasiperiodic modulation strengths addressed in Fig.~\ref{fig:delta}, specifically, $\delta=0.005$ in Fig.~\ref{fig:delta}(a,b) and $\delta=0.025$ in Fig.~\ref{fig:delta}(c,d), correspond to the situations where the array of straight waveguides is situated below and above the localization transition (see Fig.~\ref{fig:LDT}). As a result, in the array with $\delta=0.005$ the Floquet modes are delocalized for zero and small oscillation amplitude $r$, i.e., the averaged form-factors are close to zero. However, sufficiently large  $r$  leads to the existence of \textit{localized Floquet modes} with large form factors, even though the corresponding array of straight waveguides is below the localization transition. The localized Floquet modes emerge for values of $r$ around the pseudocollapses of Floquet bands in the array without quasiperiodic modulation. Hence for slow oscillations with  $\omega=0.05$ and weak quasiperiodic modulation $\delta=0.005$ there is a sequence of peaks in the distribution of averaged form factors $\tchi$, as shown in Fig.~\ref{fig:delta}(a). In Fig.~\ref{fig:floquet}(a,b) we show two representative localized Floquet modes in the array with slow oscillations, $\omega=0.05$, and with weak quasiperiodic modulation, $\delta=0.005$. The shown intensities display nearly perfect periodic propagation with no appreciable diffraction even for distances equal to ten oscillation periods, $z/Z \sim 10$. For fast longitudinal oscillations with $\omega=0.2$, there is a single peak in the  distribution of averaged form-factors within the considered interval of oscillation amplitudes $r$, see Fig.~\ref{fig:delta}(b). 

\begin{figure}
	\begin{center}
		\includegraphics[width=0.999\columnwidth]{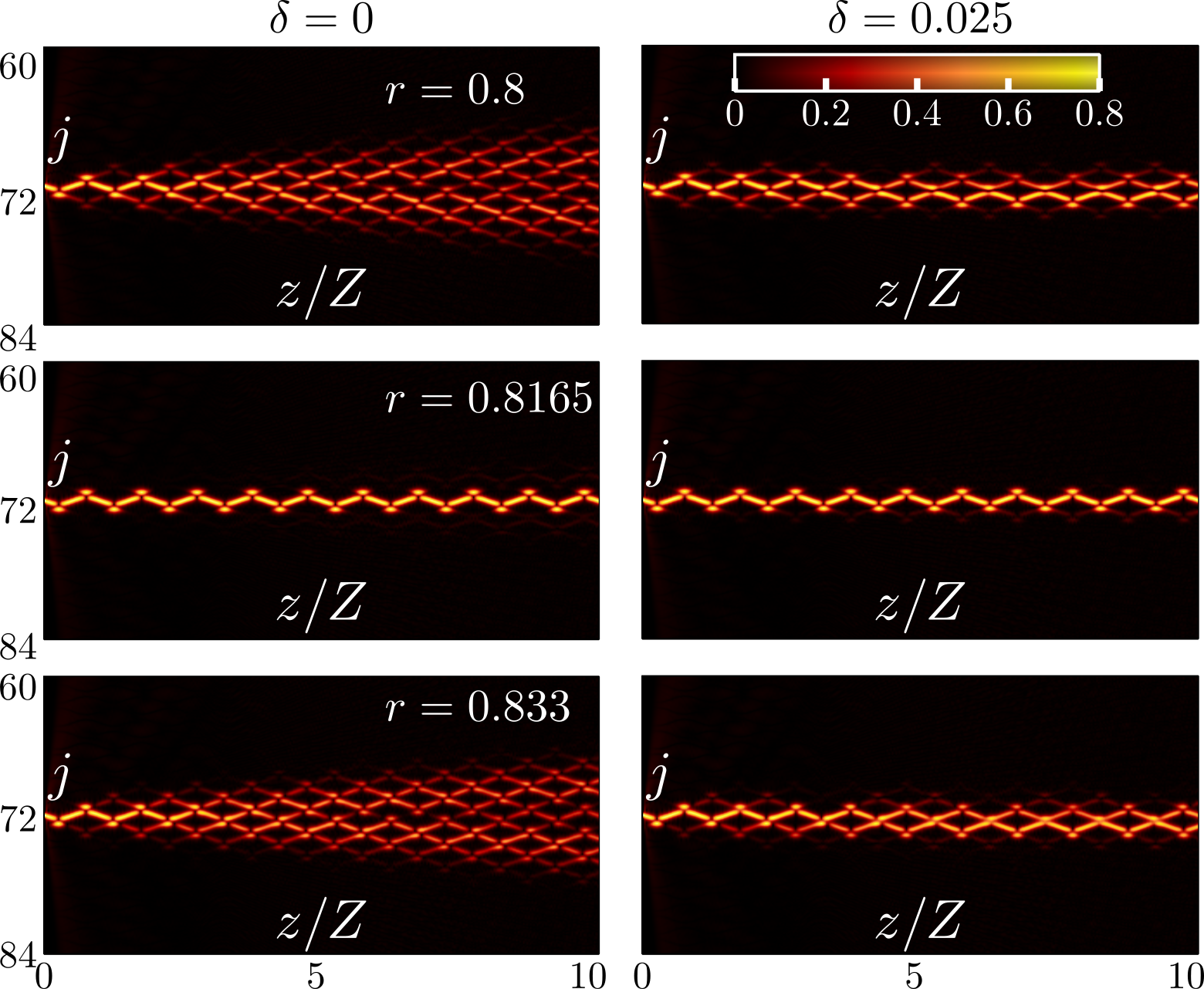}
	\end{center}
\caption{Propagation of a single-channel initial excitation in oscillating waveguide array without (left panels) and with (right panels) transverse quasiperiodic refractive index modulation when the amplitude of oscillations is detuned below ($r=0.8$, top row) or above ($r=0.833$, bottom row) the exact value corresponding to the pseudocollapse of the Floquet spectrum ($r=0.8165$, middle row). \rev{The colorbar shows $|\Psi|$ and applies to all plots.}}
	\label{fig:single}
\end{figure}

For the case of strong quasiperiodic  modulation, $\delta=0.025$, illustrated in Fig.~\ref{fig:delta}(c,d), the array is above the localization transition already in the absence of waveguide oscillations, i.e., at $r=0$. Hence, the Floquet eigenmodes are localized already at small values of $r$. In the meantime, for larger $r$  the Floquet eigenmodes can undergo secondary localization and delocalization transitions. In the array with zero quasiperiodic modulation, diffraction is appreciably suppressed at the value of $r$ corresponding to the pseudocollapse. Hence the single-channel initial excitation diffracts strongly if the amplitude $r$ is not precisely tuned to the value corresponding to the pseudocollapse, see plots in the left column of Fig.~\ref{fig:single}. In contrast, in quasiperiodically modulated array  the localized Floquet modes exist in continuous intervals of oscillation amplitudes $r$, see for example Fig.~\ref{fig:delta}(d), where localized Floquet modes exist within the interval $0.77 \lesssim r \lesssim 0.88$. Hence  the diffraction of the initial single-channel excitation is strongly suppressed even if the amplitude of oscillations considerably detuned from the pseudocollapse point, as illustrated in the plots in the right column of Fig.~\ref{fig:single}. 

\rev{While the localized Floquet modes stroboscopically reimage their shapes after each longitudinal period $Z$, their fields $\Psi^F(x,z)$ change  along the propagation distance [see  examples in Fig.~\ref{fig:floquet}(a,b)]. As a result, the form-factor $\tchi^{(2)}$, averaged over the period $Z$, may be slightly smaller  than the form-factor of nearly stationary localized states in the array with weakly curved waveguides, i.e., in arrays with $r \ll 1$.}

\rev{We have additionally addressed the effect of focusing nonlinearity on propagation of localized Floquet modes by simulating their evolution governed by the envelope equation (\ref{eq:main}) with an additional nonlinear term $g|\Psi|^2\Psi$, where the negative coefficient $g$ describes the nonlinearity strength. Weak nonlinearity [$g=-0.1$ in Fig.~\ref{fig:floquet}(c)] excites additional Floquet modes localized close to the initially excited one. As a result, the  initial intensity distribution slightly broadens but, nevertheless, remains localized over multiple longitudinal periods. Stronger nonlinearity [$g=-1$ in Fig.~\ref{fig:floquet}(d)] leads to the self-trapping of light practically in a single waveguide.}

\begin{figure}
	\begin{center}
		\includegraphics[width=1.0\columnwidth]{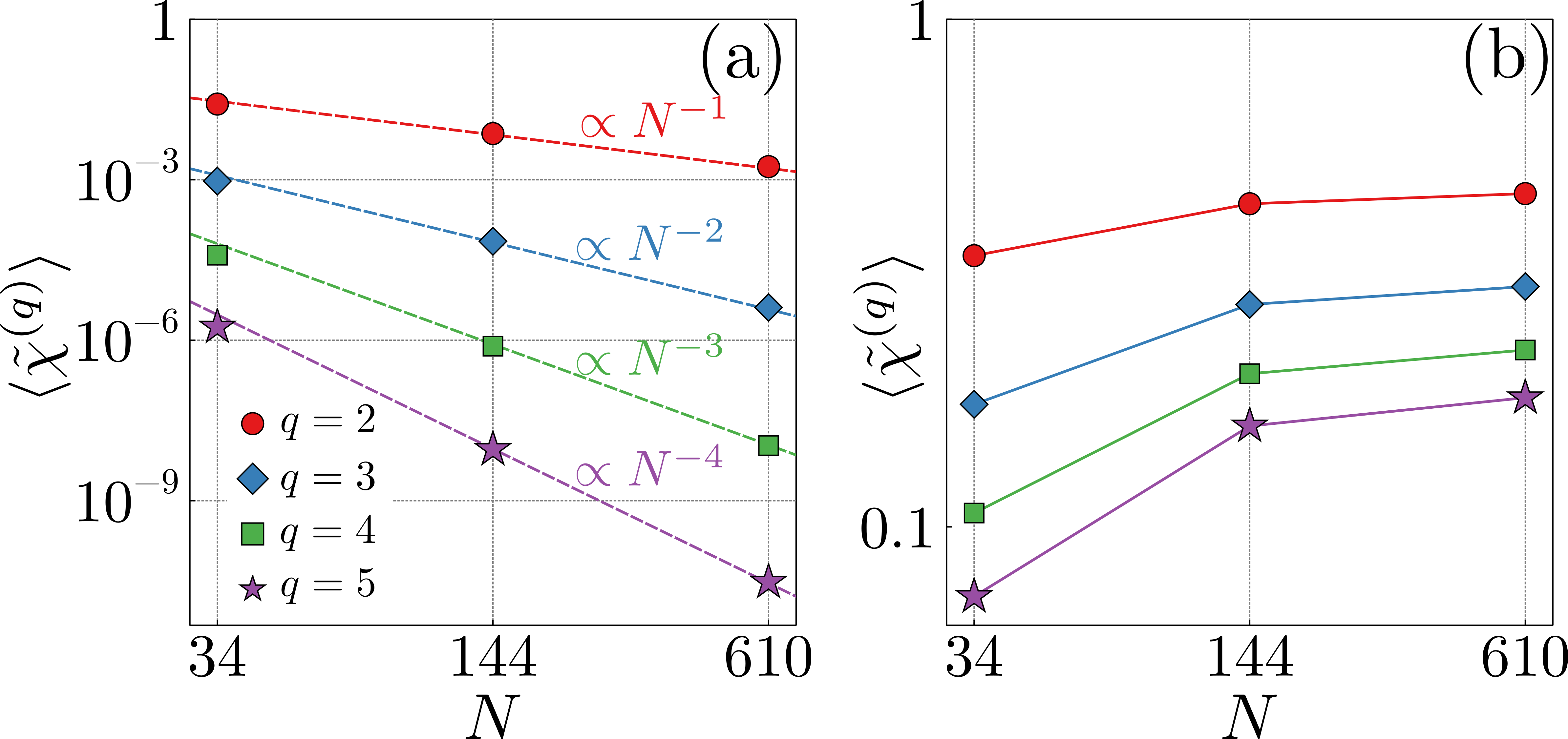}
	\end{center}
\caption{Generalized form factors of Floquet states (averaged over $N$ states) for waveguide arrays of three different length $N$ for delocalized states ($r=0.7$, a) and localized states ($r=0.815$, b) for   frequency of longitudinal modulation $\omega=0.2$ and quasiperiodic modulation strength $\delta=0.05$.  The numerical data are shown with markers. The dashed lines in (a) show linear fits. Solid lines in (b) are guide for the  eye.  }
	\label{fig:scaling}
\end{figure}

\rev{Finally, we have studied the scaling of localized and delocalized Floquet states against the number of waveguides $N$. Generalized form-factors of Floquet states $\tchi^{(q)}$  have been computed    for $N=34,\, 144,\, 610$ (which correspond to three rational approximants of increasing accuracy: $\varphi \approx 55/34,\, 233/144,\, 987/610$. Then the form factors have been averaged over all $N$ Floquet states. The obtained dependencies are presented in Fig.~\ref{fig:scaling}. For delocalized states we find the standard scaling $\propto N^{-(q-1)}$, while for localized states the averaged form-factors slightly increase with $N$ and exhibit the tendency to saturate with $N$ (compare the cases $N=144$ and $N=610$). This confirms the scaling $\propto N^0$ for localized Floquet states. }

\section{Conclusion} 
\label{sec:concl}

To conclude, we studied the competition between two localization mechanisms and introduced a previously unexplored class of localized Floquet modes whose intensity distribution is localized in the transverse direction and is periodic in the longitudinal direction. These states emerge near the pseudocollapses of Floquet band of the array with zero quasiperiodic modulation. Localized Floquet modes exist even for weak quasiperiodic modulations, when the array with straight waveguides is below the localization transition. In addition, the localized Floquet modes exist within continuous parametric intervals and can therefore be used for robust diffraction inhibition.

\rev{The preliminary results regarding the propagation of localized Floquet states in nonlinear media raise a question about the existence of truly nonlinear Floquet modes bifurcating from the linear ones, similar to $\pi$-solitons that have been recently observed  in periodic arrays of out-of-phase curved waveguides \cite{Arkhipova2023}.  }

\section*{Acknowledgments} 

The authors thank Yiqi Zhang for the assistance with the preparation of Fig.~\ref{fig:array}.  Y.V.K. acknowledges funding by the research project FFUU-2024-0003 of the Institute of Spectroscopy of the Russian Academy of Sciences.  The work of D.A.Z. was supported by the Priority 2030 Federal Academic Leadership Program.

\end{document}